# How thermal priming of coral gametes shapes fertilization success

Antoine Puisay[a,b,c], Laetitia Hédouin[a,b], Rosanne Pilon[a,b], Claire Goiran[b,d], Benoit Pujol *[a,b]

[a] CRIOBE, EPHE, Université PSL, UPVD, CNRS, UAR CRIOBE, BP1013, 98729 Moorea French Polynesia

[b] Laboratoire d'Excellence "CORAIL", 58 avenue Paul Alduy 66860 Perpignan Cedex France

[c] Present address, Coral'ite, La Jeunerie, 53290, Bierné-les-villages, France

[d] ISEA, Université de la Nouvelle-Calédonie, BP R4, 98851 Nouméa Cedex, New Caledonia

*Corresponding author: benoit.pujol@univ-perp.fr

**Highlights:** • Rising seawater temperatures reduce fertilization success in spawning corals

• We examined the understudied role of coral pelagic life stages in fertilization success

• Effect of experimental thermal pre-exposure of gametes was tested

• Coral gamete thermal priming improved fertilization success at high temperature

• Thermal priming of sperm or oocytes differentially restores fertilization success

**Graphical abstract:**

*Problem*
**Reduction of the fertilization success at high temperature**

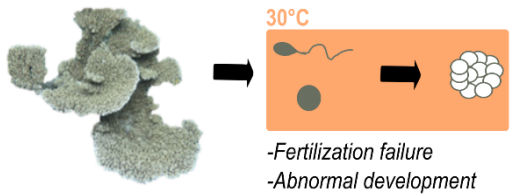

-Fertilization failure
-Abnormal development

*Question*
**Does thermal priming of coral gametes influence fertilization success?**

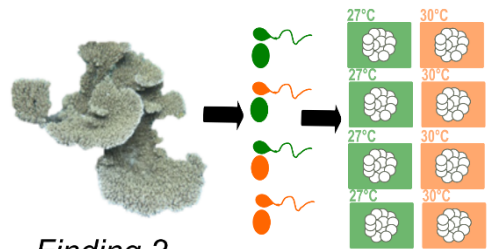

*Finding 1*
**Thermal priming of coral gametes shapes fertilization success**

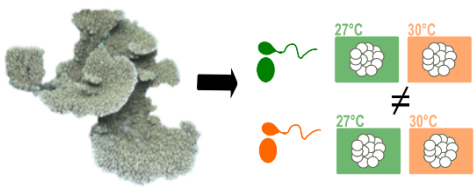

-Thermal priming of gametes can increase fertilization suceess at 30°C

*Finding 2*
**Thermal priming of sperm and oocytes does not impact fertilization success equally**

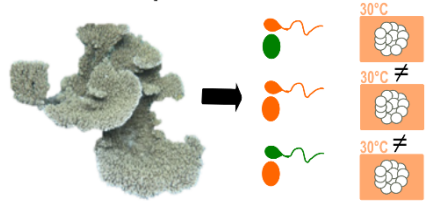

-Thermal priming of sperm is not equal to oocytes priming
-Thermal priming of both gametes showed the highest potential for acclimatization of fertilization success under high temperature

**Abstract:** Seawater temperature rise is damaging coral reef ecosystems. There is growing evidence for the negative impact of rising temperatures on the survival of adult corals and their reproductive success. However, the effect of elevated temperatures on gametes remains scarcely studied. Here we tested the effect of the thermal priming of gametes on the fertilization success in experimentally tested populations of *Acropora cytherea* corals in French Polynesia. As expected, a temperature of 30°C (ambient +3°C) reduces coral fertilization success. However, the thermal exposure of gametes to 30°C after their release in seawater prior to fertilization limited fertilization failure, with a greater impact of oocytes in comparison to sperm. This temperature is similar to temperatures observed in nature under the changing climate. Our findings imply that the thermal priming of early life stages, such as gametes may play a role in maintaining the coral fertilization success in spite of increasing seawater temperature.

1 **Introduction**

According to the 2018 IPCC climate change forecast, an increase of 2.0°C by the mid-century could lead to the worldwide collapse of coral reefs, and the loss of 99% of reef-building corals (Hoegh-Guldberg et al., 2018). Corals have a life cycle composed of a benthic adult phase that can reproduce asexually through fragmentation and sexually through the fertilization of oocytes with sperm (Harrison and Wallace, 1990). Different reproductive systems can be found in corals. The majority of sexually reproducing corals are hermaphrodites rather than gonochore (either male or female) but they nevertheless spawn egg and sperm, sometimes together in bundles, for external rather than internal fertilization. In terms of larvae dispersal, brooding corals are characterized by larvae that can settle immediately whereas larvae from broadcast spawning corals have a pelagic developmental phase that can last weeks before settlement (Harrison and Wallace, 1990). During these phases, an increase in temperature can influence physiological responses. During the adult benthic phase, coral bleaching – the loss of their obligate photo-symbiont (i.e. family Symbiodiniaceae in LaJeunesse et al., 2018) – is widely recognized as a major cause of mortality. Coral bleaching and mortality is triggered by exposure to higher temperatures than average maximal summer temperatures, in particular when they persist for a long period of time reaching several weeks (Hughes et al., 2018; Kayanne, 2017; Lesser, 2011). During the pelagic phase, laboratory studies, mostly conducted on *Acropora* species shed light on issues related to higher seawater temperatures, e.g., increased reproductive failure, development rates of embryos, and abnormal development (Bassim et al., 2002; Bassim and Sammarco, 2003; Negri et al., 2007; Randall and Szmant, 2009).

Ecological memory - the ability of an ecosystem to be influenced by past events - has the potential to modify coral bleaching responses to cumulative warming episodes during the adult benthic phase of corals (Ainsworth et al., 2016; Hughes et al., 2019; Maynard et al., 2008; Penin et al., 2013; Safaie et al., 2018). Since 2015 and the emergence of the term "assisted evolution" (van Oppen et al., 2015) several studies investigated the impact of priming in corals. These studies found that priming adult corals by exposing them to higher temperatures during their brooding phase had a positive impact on the response of larvae exposed to higher temperatures (Putnam and Gates, 2015, Jiang et al. 2023, but see McRae et al. 2021 for a negative impact in *Pocillopora acuta*, and Bellworthy et al., 2019 for weak carry-over effecst associated with a bet-hedging strategy in *Stylophora pistillata*). Hormetic conditioning or environmental priming were identified as potential mechanisms explaining developmental acclimation in reef-building corals (Putnam et al., 2020). Finally, a recent review highlighted the potential genetic and non-genetic mechanisms underlying coral acclimatization (Putnam, 2021). This review outlined the importance of investigating environmental mechanisms affecting gamete life stages and fertilization because they may influence coral acclimatization by determining offspring ecological performance. However, to the best of our knowledge, only one publication to date has specifically tested carryover effects during early life stages in a coral broadcast-spawner (Puisay et al., 2018). In terms of acclimatization to higher temperatures, these carryover effects might be beneficial or maladaptive. Temperature effects on gametes might be long-lasting because they have the potential to extend to subsequent life stages (Podolsky and Moran, 2006), notably in marine organisms with a life cycle like corals. Assessing the effect of the thermal exposure of coral gametes (intensity and time) after their release in seawater, on the fertilization success rate, is critical to get a complete picture of biological mechanisms driven by climate change that impact fitness-relevant functions (Adriaenssens et al., 2012). Furthermore, such knowledge has implications for examining the occurrence of carryover effects from one life history stage to another and for our understanding of the acclimatization potential of corals to rising seawater temperatures.

Acclimatization opportunities for corals may exist that rely on genetic and environmental mechanisms driven by higher seawater temperatures and their impact on gametes after their release in seawater. For example, an experimental study conducted on the coral *Acropora pulchra*, showed that exposing gametes at 32°C before fertilization had the potential to enhance fertilization success when fertilization occurs at 32°C (Puisay et al., 2018). Whether sperm and oocytes play a similar role in this thermal priming (acquired stress tolerance through exposure to non-lethal stress) acclimatization mechanism remains however unknown. Environmental parameters may influence differentially the contribution of maternal and paternal gametes to embryo formation and development. Identifying individual gamete responses (sperm versus oocyte) to different temperatures might enhance our ability to understand how temperature shapes coral reproductive success.

Here we tested experimentally for the hypothesis that the temperature experienced by coral male and female gametes after their release in seawater, but before fertilization, differentially shapes fertilization outcomes (e.g., fertilization success). We conducted this experiment on gametes of *Acropora cytherea* (Dana, 1846), collected from fragments originated from gravid coral colonies sampled on the fringing reef (1-1.5 m depth) off the coast of Mo'orea. Mo'orea island suffered at least seven bleaching events from 1979 to 2007, with four well documented events in 1991, 1994, 2002 and 2007 (Penin et al., 2007; Pratchett et al., 2013). Since 2007, no massive bleaching event was recorded up to the bleaching event that occurred in 2016. While the corals used in our experiment may have been impacted by these bleaching events, their last seven years were not punctuated by episodes of major heat stress. At the Center of Insular Research and Observatory of Environment (CRIOBE) in Mo'orea, French Polynesia, we separately exposed a combination of oocytes and sperm of seven colonies after their release in seawater to either 27 or 30°C during four periods of time (0.5, 1, 1.5, 2 h), before fertilization was achieved at either 27 or 30°C.

## 2 Material and methods

*2.1 Collection of corals and gametes*

We identified gravid coral colonies (oocytes appear pink in internal branches) of the simultaneous hermaphrodite broadcast-spawner *Acropora cytherea* (Dana, 1846) on the north coast of Mo'orea (on a reef named Papeto'ai), which is the most populated area of the island. We fragmented (25x25 cm) ten of these coral colonies (of an approximate size of 50x80 cm) using hammer and chisel (with a minimum distance of 10 m between colonies), two days before the full moon of the expected spawning month (Carroll et al. 2006) and transported them on a boat using wet towels in an estimated time of 20 minutes, which has proven an efficient method that protects the samples from unwanted stress caused by motion in bags or containers. The thermal regime of the spawning colonies on the reef during a year before the time of spawning was characterized by temperatures ranging between 26°C and 30°C, and between 27°C and 28.5°C two weeks before the time of sampling (Figure S2). Coral fragments were maintained in an open circuit aquarium (3000 L; T°C : 27.3 ± 0.5°C, salinity: 35; pH: 8.1 ± 0.1), with a natural 12 h/12 h dark / light cycle with a Photosynthetic Active Radiation spectral range recorded between 500 to 750 nanometers. Observation Service CORAIL from CRIOBE kindly provided the ecological monitoring data. They were monitored every-night until the spawning occurred in October 2014, seven days After Full Moon (16/10/2014, 10:08 p.m.). Spawning activity was monitored by assessing polyp gravidity, with polyps being inflated and pink during the setting phase of egg-sperm bundles that is typical of acroporid coral reproduction. At the stage when pink oocytes of egg-sperm bundles aggregate at the mouth of the polyp, gravid colonies were isolated in individual tanks. The release of gametes by corals occurred within approximately 30 min for all colonies. Gametes were collected by pipetting egg-sperm bundles into a 50 mL tube (one per colony) and were stored individually (in a 50 mL Falcon tube) for each spawning colony (n=7). To avoid the dilution of the sperm, egg-sperm bundles were concentrated with a ratio of 5 mL of bundles to 5 mL of seawater. Thus, colonies that did not spawn enough bundles (n=3) to reach that concentration were not selected for the fertilization trial of that day, but where kept at the lab for further experimentation. Eggs do not begin hydrating and sperm does not activate swimming motion until after the bundles rupture, which can happen quickly after bundle release. If some gametes are released at different times in the same pool, or bundles broke apart at different times, a potential effect of gamete age – the time between release and fertilization – might have induced some random noise in the data but we do not expect a bias. This is because bundles were broken apart by gentle agitation of the tube within 45 min after collection. The oocytes and sperm from each tube were then filtered through a plankton mesh (100 µm), which retained oocytes but not sperm. Oocytes, isolated on the mesh were rinsed five times using filtered (0.7 µm) sperm-free seawater and stored individually (in a bucket with a 100 µm mesh and 100 mL of sperm-free seawater). Sperm from all spawning colonies (first stored in the tube used for the collection of bundles) were pooled in a tube (in a 50 mL Falcon tube), with an equal amount of sperm from each colony. While the sperm concentration, as well as the sperm motility were not assessed for each colony, our protocol allowed us not to expect that heterogeneity in these parameters would bias our results. The separation of bundles, their cleaning and the total concentration estimate of the mixed sperm solution was done in 45 min. The concentration of the sperm solution (mixed of all colonies) was estimated by using a hemacytometer. The final sperm concentration (mixed of all colonies) used for the experiments was defined at $10^6$ sperm.mL$^{-1}$, as to ensure a high fertilization success (>80%) (Nozawa et al., 2015). Oocytes were pooled few minutes before the start of the experiment to avoid unwanted fertilization due to potentially remaining sperm attached to oocytes. Pooling oocytes was necessary to keep the time schedule of the experiment but made it impossible to exclude that some genotypes fertilized more eggs than others. This issue challenges the generalization of our findings but does not challenge experimental proof of concept evidence for pre-exposure effects investigated here. Overall, we controlled the time from spawning to fertilization for all gametes from all colonies, which was set to two hours, at launching of all the experiments.

*2.2 Thermal exposure of gametes*

In order to test for the role of the thermal priming of coral gametes after their release in seawater on their subsequent fertilization success, we considered both male and female gametes independently. The thermal priming of sperm and oocytes was achieved by exposing them to different temperatures after release but before fertilization. Gamete differential exposure to temperature also included an assessment of the impact of the duration of the thermal exposure to test for the hypothesis that a longer exposure to elevated temperature might induce different levels of response. Sperm (S) and oocytes (O) were exposed separately for 2.0, 1.5, 1.0, and 0.5 h at 30°C before the fertilization experiment while the control was kept at 27°C for the same period. Two hours before fertilization, we first launched the 2 h treatment, then started a new treatment on a separated sample every 30 min before fertilization. The fertilization experiment was launched directly after the thermal priming treatment. Independent fertilization trials were used to combine gametes exposed to the different thermal priming treatments. These fertilization trials were conducted at two different temperatures: 27 and 30°C for 4.5 h. A total of 4 different types of crosses (n=6 replicates per temperature treatment) were performed: $S^{27}O^{27}$, $S^{30}O^{27}$, $S^{27}O^{30}$, $S^{30}O^{30}$ (with S =sperm, O = oocyte and $^{number}$ = temperature), and assessed for their fertilization success at 27 and 30°C.

*2.3 Fertilization success measurement*

Oocytes (~200 oocytes per vials) and sperm from each colony were pooled together to cross-fertilize (less than <1% of self-fertilization generally occurs in *A. cytherea* (Ramírez-Portilla et al., 2022). Vials filled with 10 mL of 0.7 μm filtered seawater (n=6 replicates per temperature treatment), and sperm adjusted to $10^6$ sperm mL$^{-1}$ were kept in seawater baths at two nominal temperatures (27, and 30°C) regulated by a Hobby® Biotherm Pro. Glass vials were left undisturbed for 4.5 h, to obtain the highest possible fertilization success and embryonic development rate (>80%) until morula stage (Oliver and Babcock, 1992). Fertilization and embryonic development processes stopped after the elapsed period by transferring embryos newly formed in 1.5 mL tubes containing seawater-formaldehyde solution (7%). After fixation (24 h) embryos translocated and preserved in ethanol (70%) for 48 h were then imaged using a binocular microscope (Leica® M80, Leica Microsystems Inc, Buffalo grove, IL, USA). The proportion of successfully fertilized (2 cells, 4 cells, > 4 cells) was determined to establish fertilization success with the least possible error by using the picture software Image J (Collins, 2007), for each vial (see Puisay et al., 2015), but this level of detail was not used in the statistical analyses.

*2.4 Calculation*

We estimated the effect of the thermal priming of male and female gametes on the fertilization success by using a Generalized Linear Model (GLM) approach. We tested and quantified the fixed effects of fertilization temperature (27 *vs* 30°C), pre-exposure time (0.5, 1.0, 1.5, 2.0 h as a continuous variable), 27 and 30°C pre-exposure temperatures for oocytes (O27 and O30) and sperm (S27 and S30), and their interactions. The normality of the data using the Shapiro test was tested, and then negative transformed by the arcsine of the square root. The statistical software R (R Core Team, 2015), was used for graphical and statistical analyses. We used the R "stats" package and the "glm" function to conduct these analyses. To estimate effects on the observed data scale, we back transformed the effects estimated on a latent scale by using the package GGally (extension from ggplot2 to plot model coefficients; Wickham, 2016). Finally, we used the package MuMIn (Multi Model Inference; Barton, 2009) to estimate goodness of fit parameters that we used to calculate *Pseudo*-R². The global model included a 3-way interaction. We decomposed the effect of this interaction into simpler effects by running two complementary GLMs (Figure S1) restricted respectively to each fertilization temperature (27 or 30°C).

## 3 Results and discussion

Results of the GLM analysis confirmed that the fertilization temperature explained a large part (pseudo-$R^2$=37%, Table S1) of the direct changes in fertilization success, from 91 ± 5% at 27°C to 75 ± 12% at 30°C (Figure 1, Table S1).

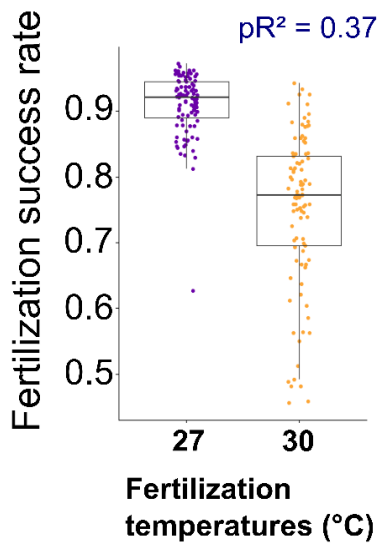

**Figure 1 Fertilization success rates in the coral *A. cytherea* are affected by fertilization temperature.**

Fertilization success rates are presented as a function of fertilization temperatures (n = 96). Effects presented here were back transformed on the real data scale (Odds-ratio). Original parameter estimates from the GLM analysis were expressed on the latent scale (Table S1). *Pseudo*-R² and other goodness of fit parameters are presented in Table S2.

Our results showed that this reduction in fertilization success observed at 30°C in comparison to 27°C was mediated by the thermal priming of gametes at a 30°C temperature before fertilization (Table S1, Figures 2 and 3). In fact, more than half of the variation in the fertilization success (pseudo-$R^2$=56%, Table S1) was significantly shaped by the combined effects of experimental treatments, as illustrated by the statistical significance of 3-way interactions (Table S1). Indeed, the different temperatures to which sperm and oocytes were exposed and the duration of this exposure could interact differentially to affect the fertilization success, depending on the fertilization temperature. Parameter estimates calculated for the 3-way interaction revealed that the fertilization success at 30°C was increased by up to 93%, all else being equal, when both sperm and oocytes were exposed to a temperature of 30°C before fertilization ($P_{Fertilization\ temperature*Time*O30xS30}$<0.001, Figures 2 and 3; 30°C:Time:O30xS30, Table S1). In comparison, this fertilization success at 30°C was increased by 64% and 62% respectively, when thermal priming was applied only to oocytes (Figures 2 and 3; 30°C:Time:O30xS27, Table S1) and sperm (Figures 2 and 3; 30°C:Time:O27xS30, Table S1). Our results therefore imply that the combined action of three conditions (i.e., gamete exposure duration and temperature, and fertilization temperature), rather than their isolated effects, critically influenced the fertilization success.

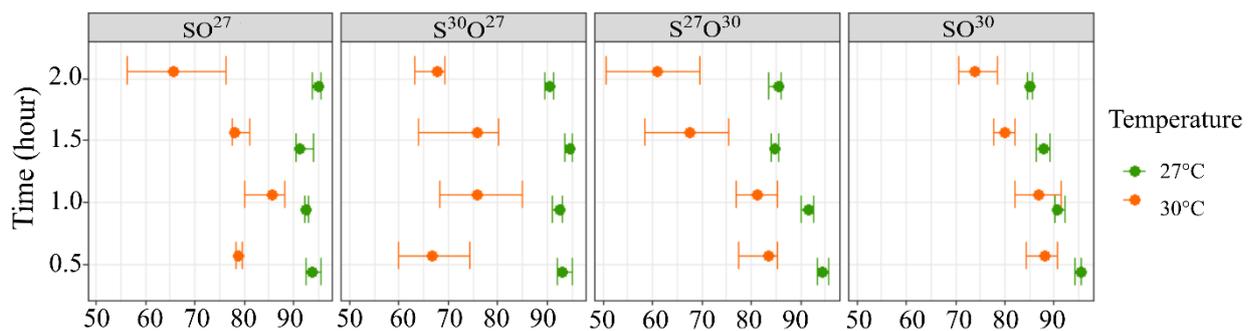

**Figure 2 Variability of the fertilization success rate**

Fertilization success median (dot) and first and third quartiles (bars), are presented on the horizontal axes for both fertilization temperatures: 27°C (green) and 30°C (orange) after gamete thermal priming. Time: duration of the pre-exposure to a specific temperature is presented on the vertical axis. The four boxes correspond to different pre-exposure temperatures: SO27: pre-exposure of sperm and oocytes to a 27°C temperature. S30O27: pre-exposure of sperm to a 30°C temperature and oocytes to a 27°C temperature. S27O30: pre-exposure of sperm to a 27°C temperature and oocytes to a 30°C temperature. SO30: pre-exposure of sperm and oocytes to a 30°C temperature.

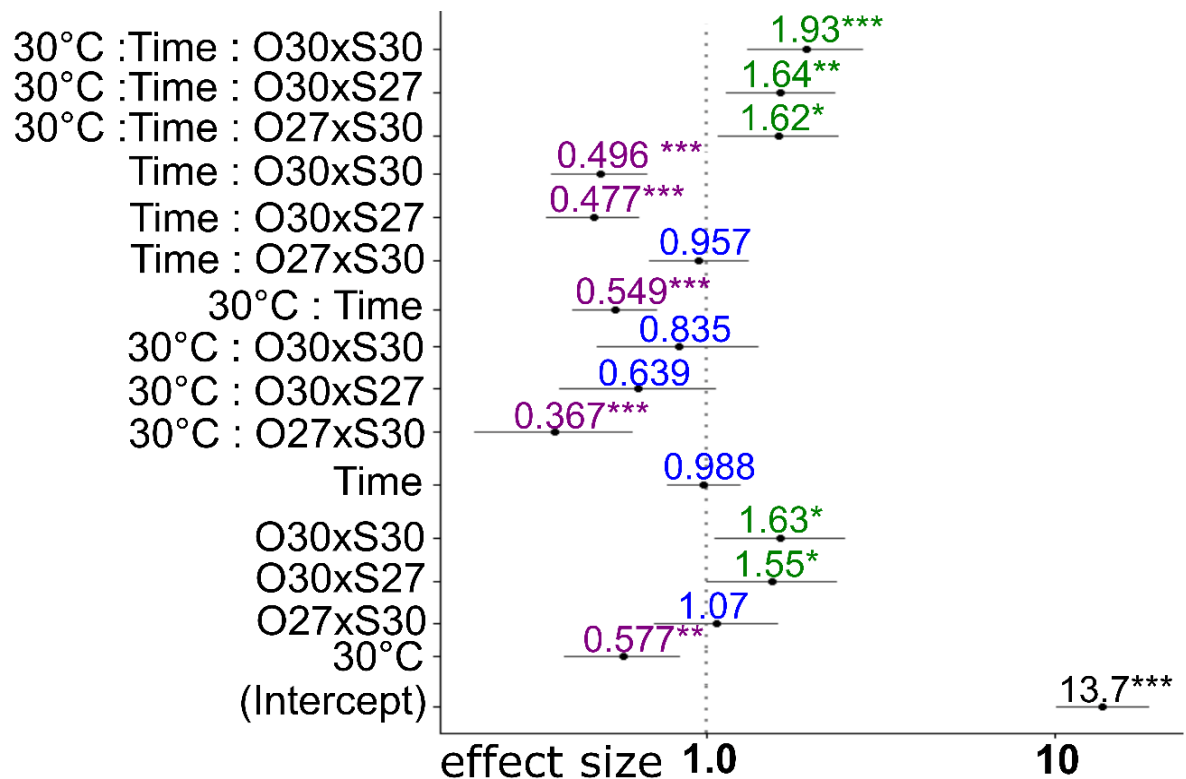

**Figure 3 Pre-exposure of female gametes to 30°C counteracts the reduction of fertilization success rates at 30°C in the coral *A. cytherea*.**

The different variables and their interactions, which effect on the fertilization success was tested, are presented on the Y axis legend: Time: duration of exposure. O30: exposure of oocytes to 30°C. S30: exposure of sperm to 30°C. O27: pre-exposure of oocytes to 27°C. S27: pre-exposure of sperm to 27°C. 30°C: Fertilization at 30°C. The effect sizes represented by their effect parameter estimates are represented on the X axis the for each variable and their interactions. For a parameter estimate to overlap 1 on the scale of effect sizes means an absence of effect of this variable. Statistical significance is presented near the estimate: * p<0.05, ** p<0.01, *** p<0.001. For the estimates, purple values represent negative effects, blue values null effects, and green values positive effects.

In the presence of 3-way interactions, caution is needed for the interpretation of other effects. For example, although increasing the duration of the thermal priming of gametes had a positive impact when fertilization occurred at 30°C, a similar increase, but with all other things being equal, had in fact a negative effect on the fertilization success; (Figures 2 and 3; Time:O30xS30 and Time:O30xS27, 30°C:Time, Table S1). Interestingly, one will easily notice a very coherent effect on Figure 2: the thermal priming of female gametes (O30xS30 and O30xS27) had a different impact on the fertilization success than the thermal priming of male gametes (O27xS30) in any given circumstance, in main factor effects or their interactions. We can therefore reliably infer that oocytes and sperm play different roles in the mediation of the negative impact of a higher fertilization temperature on the fertilization success. Furthermore, the duration of the exposure of either sperm or oocytes to 30°C before fertilization differentially affect the fertilization success. The deleterious effect of fertilization at 30°C was mediated more strongly by the thermal priming of oocytes released in seawater to a temperature of 30°C before fertilization, as illustrated by the 63% and the 55% increase in fertilization success (Figures 2 and 3; O30xS30 and O30xS27, Table S1) compared to no effect linked to the exposure of sperm to a 30°C temperature (Figures 2 and 3 O27xS30, Table S1).

The thermal environments experienced by sperm and oocytes had an impact on the fertilization success outcome. This impact could result from the trial of some gametes better adapted to higher temperatures during the pre-exposure treatment. As a result, mating probability and fertilization success would therefore likely be higher for those gametes. Such selection mechanism, which would imply the presence of genetic variation for mating performance amongst gametes, remains to be tested. Environmental effects such as the temperature encountered during pre-exposure treatments may also have directly affected (e.g., phenotypic plasticity) gametes. These two mechanisms – selection and plasticity – are not exclusive. Previous findings implied that the reduction of the fertilization observed at a 30°C temperature may be due to a reduction of the number of oocytes fertilized, rather than an increase in the abnormal development of embryos (Puisay et al., 2018). Abnormal development may result from the environmental temperature during embryonic development, rather than before. For instance, increased metabolism influenced the swimming speed of sperm in marine invertebrates (Kupriyanova and Havenhand, 2005) and the thermal tolerance of oocytes in corals (Puisay et al., 2018). Potential biological mechanisms underlying our findings might be increased metabolism and oxygen consumption, as previously observed for gametes in warming environments (Kupriyanova and Havenhand, 2005).

Interestingly, all things being equal, the fertilization success was more variable at 30°C, which may indicate opportunity for selection during the fertilization success. Furthermore, when considering the particular treatment where the thermal priming of gametes increased the fertilization success at 30°C, we observed a reduction of the variation of the fertilization success (Figure 2). This experimental reduction in the variation of the fertilization success may imply that selection occurred on a pool of diverse gametes, which translated into a better resilience to 30°C but ultimately would result in a reduction of adaptive potential by narrowing down the potential targets for future selection. Caution must be taken before concluding that fertilization success variability corresponds to potential for adaptive evolution in response to increased seawater temperature because there is no evidence for its genetic basis or transgenerational plasticity. A screening of the genetic and metabolic differences among different temperature

16treatments would give more insight into the interaction of genetic and environmental mechanisms in early life stages, and the occurrence of carryover effects and their adaptive significance. Furthermore, more work is needed to identify the physiological mechanisms underlying our findings and understanding how the pre-exposure effects that we identified affect populations in the wild.

Our findings are consistent with the mechanisms of coral acclimatization generally observed in corals (Putnam, 2021). Our results showed that the thermal priming of either oocytes or sperm (O27xS30, O30xS27) has the potential to reduce the reduction of fertilization success observed at 30°C, and that thermal priming of both gametes at 30°C (O30xS30) was more effective. This finding could be the outcome of a conditioning mismatch between gametes, which could be caused by epigenetic asymmetry or incompatibility (Putnam, 2021). Another interesting feature of our finding is that the thermal priming response of coral gametes appeared to follow a dose-dependent mechanism. The combination of time and temperature influenced positively the fertilization success rate, which might be linked to hormesis (Hackerott et al., 2021; Putnam et al., 2020). Our findings also suggest that the positive impact of the duration of thermal priming may vary in corals depending on the life stage, cell activity and cell type encountered, with more time required, which is counted in weeks, for adult corals (Hughes et al., 2019; Pratchett et al., 2013) than for larvae, where thermal priming time was counted in days (Ross et al., 2013) or gametes, as illustrated by our study where thermal priming time was counted in hours.

**Conclusion**

Our findings suggest that gamete thermal exposure treatments before fertilization and in particular 0.5 h and 1.0 h at 30°C maximizes *A. cytherea* chances of successful fertilization in the context of experimental rearing. Our experiments on *A. cytherea* corals showed that thermal priming influenced up to 56% of the reproductive success observed in a warmer (+3°C) and therefore deleterious environment. This is probably resulting from the increased tolerance of gametes and the selection of the best gametes in response to warming conditions. Our findings have implication for fundamental research, as they suggest the thermal tolerance and possibly the acclimatization of gametes, with differential effects on sperm and oocytes. Although coral gametes biology and their thermal histories are underrepresented in the scientific literature, this often neglected section of the life cycle of corals critically influences their reproductive success. Our findings imply that findings on the biology of gametes must be integrated with results on other life stages, in particular as carryover effects might have latent effects on successive life stages (i.e., larvae, recruits, and adults).

**Acknowledgements**

We thank the CRIOBE staff for their help during the field collection and lab experiments. We also thank the National Center of Scientific Research and the "Ecole Pratique des Hautes Etudes" for their support. We also thank "coral'ite" (consulting freelance: R&D, Education). We thank Aurélie Moya and Hollie Putnam for their comments on a previous version of this manuscript.

**Credit author statement**

**Antoine Puisay**: Conceptualization, Data curation, Formal analysis, Investigation, Methodology, Writing - original draft, Writing - review & editing. **Rosanne Pilon**: Conceptualization, Data curation, Methodology, Writing - review & editing. **Claire Goiran**: Data curation, Writing - review & editing. **Laetitia Hédouin**: Conceptualization, Data curation, Formal analysis, Funding acquisition, Investigation, Methodology, Project administration. **Benoit Pujol**: Formal analysis, Investigation, Writing - original draft, Writing - review & editing Writing.

**Funding:**

This work was funded by the French National Research Agency under the call "RPDOC 2010" with a project entitled "R-ECOLOGS", number ANR-10-PDOC- 0013; the Labex CORAIL with a project entitled "ARCOS"; the Ministère d'Outre Mer with a project entitled "AQUA-CORAL" and the Contrat Etat-Projet 2016-18. This work was supported by a PhD grant from Association National Research and Technology in collaboration with CRIOBE laboratory and InterContinental Tahiti Resort and Spa. Funding was not involved in the design of the study collection, analysis and interpretation of the data, the writing of the report or the decision to submit the article for publication.

**Data accessibility statement:** Should the manuscript be accepted, the data supporting the results will be archived in an appropriate public repository (Zenodo) and the data DOI will be included at the end of the article. https://doi.org/10.5281/zenodo.8032394

**Conflict of interest**: Authors declare no conflict of interests.

## Supplementary information

Table S1: Results from generalized linear models on the fertilization success of the coral *Acropora cytherea* at 27 and 30°C. The model column indicates the results for separated analyses (i) at 27 and 30°C, (ii) 27°C, and (iii) 30°C. Bold variables and *p*-values represent significant tests.

|  | model | Estimate | Std.error | z value | Pr > \|z\| |
|---|---|---|---|---|---|
| **Fertilization temperature 30°C** | 27-30 | -0.54910 | 0.19302 | -2.845 | **0.004444 \*\*** |
| Time | 27-30 | -0.01191 | 0.12225 | -0.097 | 0.922419 |
| O27xS30 | 27-30 | 0.07016 | 0.21079 | 0.333 | 0.739238 |
| **O30xS27** | 27-30 | 0.44008 | 0.21956 | 2.004 | **0.045033 \*** |
| **O30xS30** | 27-30 | 0.48908 | 0.22010 | 2.222 | **0.026278 \*** |
| **Fertilization temperature 30°C x Time** | 27-30 | -0.59972 | 0.14501 | -4.136 | **3.54e-05\*\*\*** |
| **Fertilization temperature 30°C x O27xS30** | 27-30 | -1.00276 | 0.26809 | -3.740 | **0.000184\*\*\*** |
| Fertilization temperature 30°C x O30xS27 | 27-30 | -0.44829 | 0.26317 | -1.703 | 0.088496 |
| Fertilization temperature 30°C x O30xS30 | 27-30 | -0.18080 | 0.27235 | -0.664 | 0.506778 |
| Time x O27xS30 | 27-30 | -0.04372 | 0.17023 | -0.257 | 0.797315 |
| **Time x O30xS27** | 27-30 | -0.74019 | 0.15682 | -4.720 | **2.36e-06\*\*\*** |
| **Time x O30xS30** | 27-30 | -0.70096 | 0.16095 | -4.355 | **1.33e-05\*\*\*** |
| **Fertilization temperature 30°C x Time x O27xS30** | 27-30 | 0.48433 | 0.20358 | 2.379 | **0.017358\*** |
| **Fertilization temperature 30°C x Time x O30xS27** | 27-30 | 0.49569 | 0.18574 | 2.669 | **0.007616\*\*** |
| **Fertilization temperature 30°C x Time x O30xS30** | 27-30 | 0.65865 | 0.19366 | 3.401 | **0.000671\*\*\*** |
| Time | 27 | -0.01191 | 0.12225 | -0.097 | 0.9224 |
| O27xS30 | 27 | 0.07016 | 0.21079 | 0.333 | 0.7392 |
| **O30xS27** | 27 | 0.44008 | 0.21956 | 2.004 | **0.0450 \*** |
| **O30xS30** | 27 | 0.48908 | 0.22010 | 2.222 | **0.0263 \*** |
| Time x O27xS30 | 27 | -0.04372 | 0.17023 | -0.257 | 0.7973 |
| **Time x O30xS27** | 27 | -0.74019 | 0.15682 | -4.720 | **2.36e-06 \*\*\*** |
| **Time x O30xS30** | 27 | -0.70096 | 0.16095 | -4.355 | **1.33e-05 \*\*\*** |
| **Time** | 30 | -0.611623 | 0.077994 | -7.842 | **4.44e-15 \*\*\*** |
| **O27xS30** | 30 | -0.932595 | 0.165650 | -5.630 | **1.80e-08 \*\*\*** |
| O30xS27 | 30 | -0.008207 | 0.145096 | -0.057 | 0.9549 |
| O30xS30 | 30 | 0.308280 | 0.160402 | 1.922 | 0.0546 . |
| **Time x O27xS30** | 30 | 0.440606 | 0.111654 | 3.946 | **7.94e-05 \*\*\*** |
| **Time x O30xS27** | 30 | -0.244499 | 0.099539 | -2.456 | **0.0140 \*** |
| Time x O30xS30 | 30 | -0.042309 | 0.107706 | -0.393 | 0.6944 |

Table S2: *Pseudo*-R² estimates and other goodness of fit parameters for the generalized linear models, i.e., global model combining 27 and 30°C, and separated models at 27°C and 30°C.

| | model | Df | logLik | LogLik modnull | AICc | Delta AIC | weight | *Pseudo*R² |
|---|---|---|---|---|---|---|---|---|
| **Pre-exposure temperature** | 27-30 | 4 | -1734.399 | -1809.766 | 3477.012 | 1854.482098 | 0.00E+00 | 0.04164461 |
| | 27 | 4 | -320.8056 | -363.8839 | 650.0508 | 101.78046 | 7.92E-23 | 0.118384738 |
| | 30 | 4 | -703.3118 | -780.2976 | 1415.063 | 340.60188 | 1.09E-74 | 0.0986621 |
| **Pre-exposure time** | 27-30 | 2 | -1555.8329 | -1809.766 | 3115.729 | 1493.199518 | 0.00E+00 | 0.14031267 |
| | 27 | 2 | -317.2566 | -363.8839 | 638.6422 | 90.3719 | 2.38E-20 | 0.128137848 |
| | 30 | 2 | -631.675 | -780.2976 | 1267.479 | 193.0179 | 1.22E-42 | 0.190469124 |
| **Fertilization temperature** | 27-30 | 2 | -1144.1816 | -1809.766 | 2292.427 | 669.896903 | 3.18E-146 | 0.367773734 |
| **Pre-exposure temperature x Pre-exposure time x Fertilization temperature** | 27-30 | 16 | -793.7106 | -1809.766 | 1622.53 | 0 | 9.31E-01 | 0.561429157 |

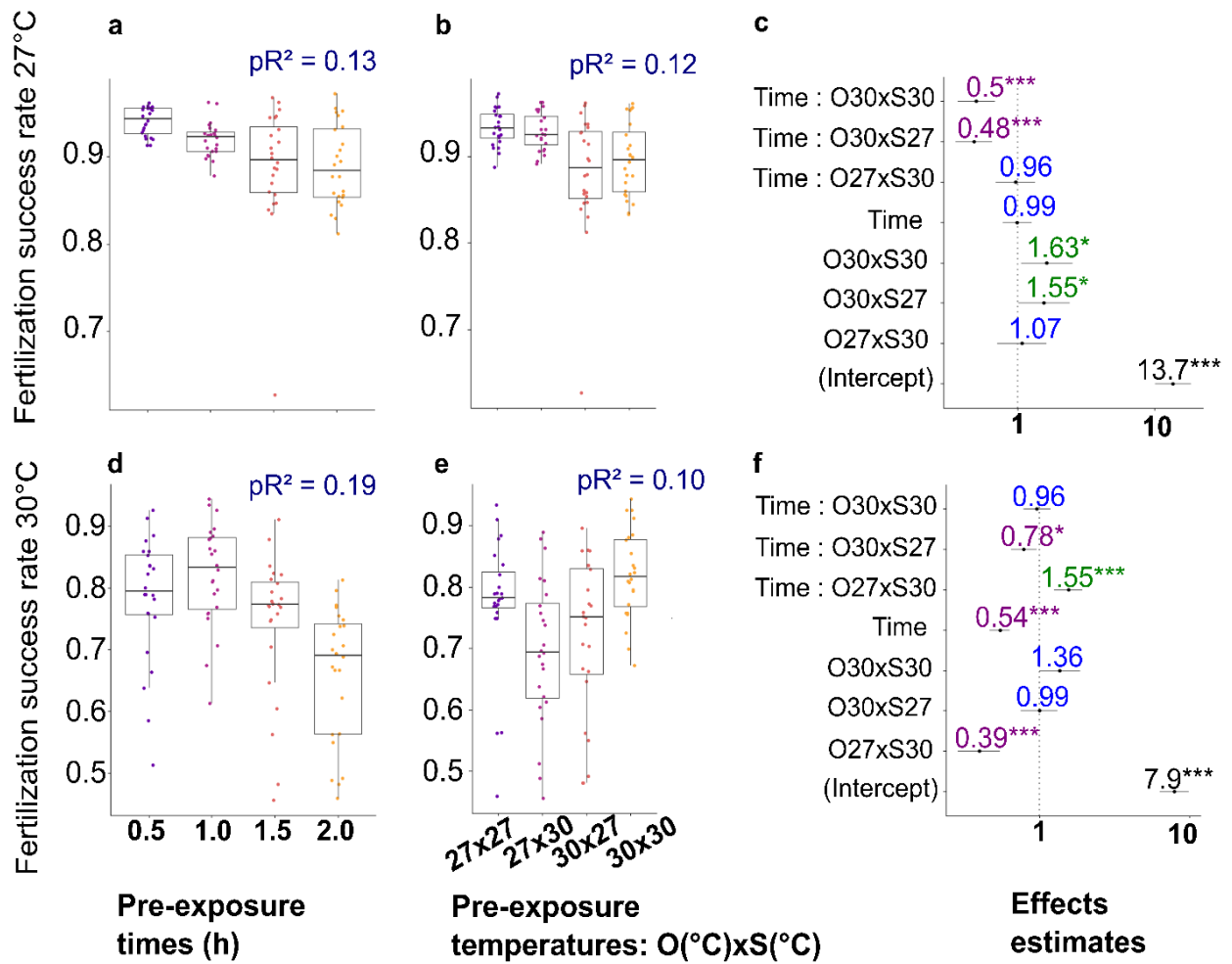

**Figure S1: Effect of pre-exposure of gametes on the fertilization success rate at 27°C and 30°C**

Effects are presented for both fertilization temperatures: 27°C (a, b, c) and 30°C (d, e, f) after back transformation on the real data scale (Odds-ratio). Original parameter estimates from the GLM analysis expressed on the latent scale can be found in Table S1. *Pseudo*-$R^2$ and other goodness of fit parameters can be found in Table S2. Effect of different pre-exposure time (a, d, n = 24), and pre-exposure temperature (b, e, n = 24), and estimates for the effects of every variable and their interactions (c, f) in the coral *A. cytherea*. Each point represents the response of an average of 160 embryos. Different colors represent different treatments (a, b, d, e). Time: duration of the pre-exposure. O30: pre-exposure of oocytes to a 30°C temperature. S30: pre-exposure of sperm to a 30°C temperature. O27: pre-exposure of oocytes to a 27°C temperature. S27: pre-exposure of sperm to a 27°C temperature. 30°C: Fertilization temperature of 30°C. For the estimates, purple values represent negative effects, blue values null effects, and green values positive effects (c, f).

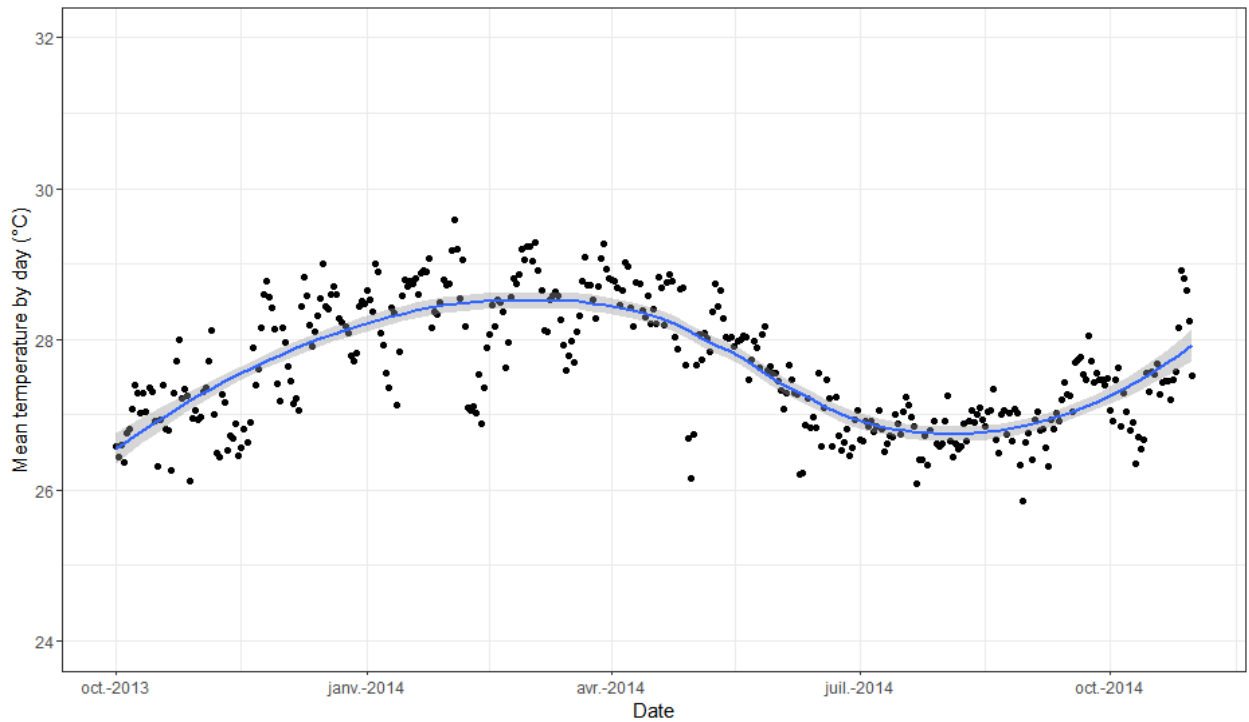

**Figure S2: Temperature profile of the corals sampling site from the first of October 2013 to the 31st October 2014.** Each dot represent the mean temperature by day. The blue line represent the smoothed temperature profile.